\pdfoutput=1

\documentclass[11pt]{article}
\usepackage{moriond,epsfig}

\def\be{\begin{equation}}
\def\ee{\end{equation}}
\def\bea{\begin{eqnarray}}
\def\eea{\end{eqnarray}}

\begin{document}

\vspace*{2cm}
\title{HIERARCHICAL MARKOVIAN ALGORITHM IN QCD EVOLUTION \footnote{Talk given at XLIIId - Rencontres de Moriond QCD and High Energy Interactions, La Thuile, Italy, March 08-15 2008.}}

\author{ A. KUSINA }

\address{Institute of Nuclear Physics, Polish Academy of Sciences,\\
	ul.\ Radzikowskiego 152, 31-342 Cracow, Poland.}

\maketitle\abstracts{
A general formalism was introduced for reorganization of QCD evolution
equations and derivation of hierarchical solution to DGLAP equation. The
hierarchical solution separates two types of parton emissions: the flavour
changing emissions and the diagonal ones (bremsstrahlung). It has been
shown that both of the achieved processes are of the Markovian type,
what gives a nice possibility of Monte Carlo implementation.
Hierarchical algorithm has been implemented and crosschecked with
standard Markovian algorithm implemented in program {\tt EvolFMC}.}

\section{Introduction}
General Form of QCD evolution equation is given by
\begin{equation}
  \label{eq:ewol}
  \partial_t D_k(t,x)=\sum_j P_{kj}(t,\cdot)\otimes D_j(t,\cdot)(x),
\end{equation}
where the convolution rule is defined by%
\begin{equation}
  f_{1}(\cdot) \otimes f_{2}(\cdot)(x) = \int d z_{1} d z_{2}
  \delta(x-z_{1}z_{2})f_{1}(z_{1})f_{2}(z_{2}).
\end{equation}
Parton momentum distribution function (PDF) $D_k$ is a function of fraction of
hadron longitudinal momentum - $x$ and evolution variable $t=\ln Q$
(where $Q$ is virtuality).
Splitting function $P_{kj}$ which represents probability of parton emission,
can be calculated perturbatively. General form of $P_{kj}$ is
\begin{equation}
  \label{eq:kernel}
  P_{kj}(t,z)=-P_{kj}^{\delta}(t)\delta_{jk}\delta(1-z) + P_{kj}^{\Theta}(t,z).
\end{equation}
We impose the momentum sum rule: $\sum_k \int_0^1 dx\, x D_k(t,x) = const$
what leads to relation between virtual and real part of kernel
\begin{equation}
P_{kk}^{\delta}(t) = \sum_j \int_0^1 dz \, z P_{jk}^{\Theta}(t,z).
\label{eq:sumrule}
\end{equation}

\section{Reorganisation of evolution equation}
Our aim is to separate two types of parton emissions: flavour changing emissions
and the diagonal ones (bremsstrahlung). To do it we divide kernel into two parts
the diagonal in flavour part and the rest
\begin{equation}
\begin{array}{rl}
  & P_{jk}^A(t,z)=-P_{jk}^{\delta\,A}(t)\delta_{jk}\delta(1-z) +
  \delta_{jk}P_{kk}^{\Theta}(t,z),\\
  & \\
  & P_{jk}^B(t,z)=P_{jk}(t,z)-P_{jk}^A(t,z)=-P_{jk}^{\delta\,B}(t)\delta_{jk}
  \delta(1-z) + (1-\delta_{jk})P_{jk}^{\Theta}(t,z).
\end{array}
\end{equation}
We do it in such a way that for both parts we have relation analogous
to equation~(\ref{eq:sumrule}).
We will use the compact matrix notation,
where $k$ and $x$ are treated as indices. But we need to remember that we do
not have the properties of commutativity anymore. Now the evolution equation has
the form:
\begin{equation}
 \partial_t {\bf D}(t)={\bf P}(t){\bf D}(t)=({\bf P}^A(t)+{\bf P}^B(t)){\bf D}(t).
\end{equation}
We introduce the diagonal transitions operator ${\bf G}_A(t,t_0)$
which is solution of the equation analogues to our evolution
equation~(\ref{eq:ewol}) but with only diagonal part of the kernel:
\begin{equation}
\partial_t {\bf G}_A(t,t_0)={\bf P}^A(t){\bf G}_A(t,t_0).
\label{eq:Ga}
\end{equation}
Solution to this equation can be written as time ordered exponent:
\begin{equation}
{\bf G}_A(t,t_0)= T \exp \bigg(\int_{t_0}^t dt^{\prime} {\bf P}^A(t^{\prime})\bigg).
\end{equation}
Than we do a change of variables $\bar{{\bf P}}^B(t) = {\bf G}_A^{-1}(t,t_0){\bf P}^B(t){\bf G}_A(t,t_0)$ and $\bar{{\bf D}}(t)={\bf G}_A^{-1}(t,t_0){\bf D}(t)$.
In this new variables evolution equation~(\ref{eq:ewol}) takes form:
\begin{equation}
\label{eq:bar}
\partial_t\bar{{\bf D}}(t)=\bar{{\bf P}}^B(t)\bar{{\bf D}}(t).
\end{equation}
Now it is easy to see that solution of equation~(\ref{eq:bar}) is given by time ordered exponent. We iterate this exponent and translate it back to original variables:
\begin{equation}
\begin{array}{rl}
  &{\bf D}(t)={\bf G}_A(t,t_0){\bf D}(t_0)+\\
  & \\
  &+\sum_{n=1}^{\infty} \bigg( \prod_{i=1}^n \int_{t_0}^t dt_i \Theta(t_i-t_{i-1})
  {\bf G}_A(t_{i+1},t_i){\bf P}^B(t_i) \bigg) {\bf G}_A(t_1,t_0){\bf D}(t_0).
\end{array}
\end{equation}
For completeness we come back to standard notation which show all the indices
explicitly. We also perform resummation of virtual part of flavour changing
kernel $P^{\delta B}$ what leads to introduction of flavour changing Sudakov
formfactor
$\Phi^B_k(t,t_0)= \int_{t_0}^t dt^{\prime}  P_{kk}^{\delta B}(t^{\prime})$.
\begin{equation}
\begin{array}{rl}
&D_k(t,x)=\int_0^1 dz^{\prime} dx_0 e^{-\Phi_k^B(t,t_0)} G^A_{kk}(t,t_0,z^{\prime}) D_k(t_0,x_0)
\delta(x-z^{\prime}x_0)\\
& \\
&+\sum_{n=1}^{\infty} \sum_{k_{n-1}\ne...\ne k_1 \ne k_0} \bigg( \prod_{i=1}^n \int_{t_0}^t dt_i 
\Theta(t_i-t_{i-1}) \int_0^1 dz_i^{\prime} \int_0^1 dz_i e^{-\Phi_{k_{i-1}}^B(t_i,t_{i-1})}
G^A_{k_{i-1}k_{i-1}}(t_i,t_{i-1},z_i^{\prime})\\
& \\
&\times  P_{k_i k_{i-1}}^{\Theta}(t_i,z_i) \bigg)
\int_0^1 dz_{n+1}^{\prime} e^{-\Phi_k^B(t,t_n)} G^A_{kk}(t,t_n,z_{n+1}^{\prime}) \int_0^1 dx_0 D_{k_0}(t_0,x_0)
\delta\big(x-x_0 \prod_{i=1}^n z_i \prod_{i=1}^{n+1}z_i^{\prime}\big),
\end{array}
\end{equation}
Now to have the full solution we need to find the explicit form of function
$G_{kk}^A(t,t_0,z)$ which represents diagonal transitions.
By definition it is a solution of equation~(\ref{eq:bar}) so we can use
analogues procedure to solve it. The solution in iterative form is given by:
\begin{equation}
\begin{array}{rl}
&G^A_{kk}(t,t_0,z^{\prime})=e^{-\Phi_k^A(t,t_0)} \delta(1-z^{\prime})+\\
& \\
&+ \sum_{n=1}^{\infty}\bigg[ \prod_{i=1}^n \int_{t_0}^t dt_i \Theta(t_i-t_{i-1})
\int_0^1 dz_i^{\prime} \bigg] e^{-\Phi_k^A(t,t_n)} \bigg[ \prod_{i=1}^n
e^{-\Phi_k^A(t_i,t_{i-1})} P_{kk}^{\Theta}(t_i,z_i^{\prime}) \bigg] 
\delta(z^{\prime}-\prod_{i=1}^n z_i^{\prime}),
\end{array}
\end{equation}
where the initial condition for $G_{kk}^A(t_0,t_0,z^{\prime})$ as Dirac delta
in $z^{\prime}=1$ was used.\\
In this way we obtained hierarchy of two processes. The {\em external} - flavour
changing process and {\em internal} - diagonal process. For both processes one
can define properly normalised transition probability which depends only on
previous step (emission).
Probability in the flavour changing process is given by:
\begin{equation}
\label{eq:step_up}
\omega(t_i,z_iz_i^{\prime}x_{i-1},k_i|t_{i-1},x_{i-1},k_{i-1})=(1-\delta_{k_ik_{i-1}})
e^{-\Phi^B_k(t_i,t_{i-1})} P_{k_i k_{i-1}}^{\Theta}(z_i)
z_i^{\prime} G^A_{k_{i-1}k_{i-1}}(t_i,t_{i-1},z_i^{\prime}),
\end{equation}
with normalisation condition:
\begin{equation}
\int_{t_{i-1}}^{\infty} dt_i \sum_{k_i} \int_0^1 dz_i \int_0^1 dz_i^{\prime} \;
\omega(t_i,z_iz_i^{\prime} \; x_{i-1},k_i|t_{i-1},x_{i-1},k_{i-1}) \equiv 1.
\end{equation}
For bremsstrahlung process we have:
\begin{equation}
p_k(t_i,Z_iz_{i-1}^{\prime}|t_{i-1},z_{i-1}^{\prime}) = \Theta(t_i-t_{i-1}) 
P_{kk}^{\Theta}(t_i,Z_i) e^{-\Phi^A_k(t_i,t_{i-1})}
\label{eq:krok_zprime}
\end{equation}
and normalisation condition:
\begin{equation}
\int_{t_{i-1}}^{\infty} dt_i \int_0^1 dZ_i \;
p_k(t_i,Z_iz_{i-1}^{\prime}|t_{i-1},z_{i-1}^{\prime}) \equiv 1.
\end{equation}
As we can see each of the processes on its own is of Markov type. This
means one can use standard algorithm to solve each of the two processes
separately. What connects this two levels are limits for evolution variable.
So by means of presented reorganisation~\cite{hier}
we obtained hierarchy of two Markov processes which has straightforward
interpretation/implementation as a generalisation of well known Markovian
algorithm.

\section{Results}
\begin{figure}[h]
\centering
\includegraphics[angle=90,width=17cm]{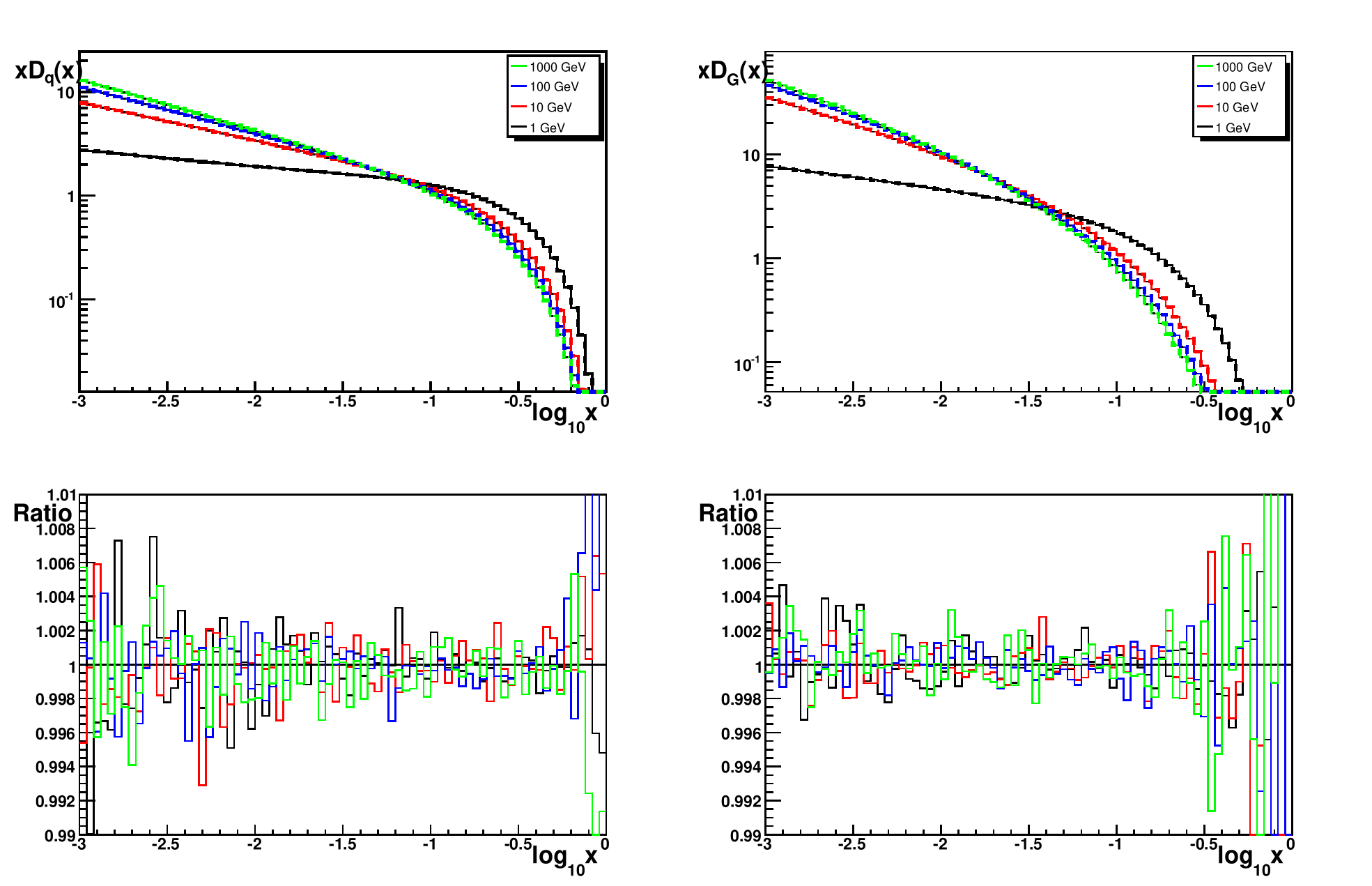}
\caption{\it The left upper plot presents gluon distributions and right upper plot quark distributions obtained from hierarchical algorithm (dashed line) and standard Markovian algorithm-{\tt EvolFMC} (straight line) for scales of 1Gev, 10GeV, 100GeV, 1000GeV. The lower plots represent coresponding ratio between this two distributions.}
\label{fig:all}
\end{figure}

Presented algorithm has been used to solve LO DGLAP evolution equation with
non-running $\alpha_S$. It takes into account three massless quarks and gluons.
PDFs were calculated for scales reaching from $Q=1$GeV (initial conditions)
up to scale of $Q=1000$GeV.
Results have been compared with standard Markovian algorithm implemented
in program {\tt EvolFMC}~\cite{MMC2006,MMC2004}.
Fig.~\ref{fig:all} shows gluon momentum distribution $xD_G(t,x)$ and quark
momentum distribution $xD_q(t,x)$ obtained from both algorithms and
ratio for coresponding distributions for scales equal 1GeV, 10GeV, 100GeV
and 1000GeV.
Presented results were calculated for statistics equal $2*10^8$ and as we can
see agreement of both approaches is on the level of $0.5$\%.

\section{Summary}
The general formalism for reorganization of evolution equation has been
used for derivation of hierarchical solution to DGLAP equation.
Very good agreement with standard Markovian algorithm
shows that it can compete with it.
Generalisation for NLO and for running $\alpha_S$ case can be done and is
only a matter of implementation. Also generalisation for different types of
evolution equations is possible. Especially case of CCFM evolution seems to
be worth considering because of possibility of simple changing of coupling
constants argument for flavour changing and diagonal emissions.

\section*{Acknowledgments}
I would like to thank S.~Jadach for useful comments and discussions.
The project is partly supported by EU grant MTKD-CT-2004-510126.


\section*{References}

\end{document}